\documentclass[fleqn,12pt]{wlscirep}
\usepackage{wrapfig, framed, caption}
\usepackage[utf8]{inputenc}
\usepackage{heuristica}
\usepackage{graphicx}
\usepackage{wrapfig, framed, caption}
\usepackage{tcolorbox}
\usepackage{xr}
\usepackage{hyperref}
\usepackage{xcolor}
\usepackage{subcaption}
\usepackage{float}

\makeatletter
\newcommand*{\addFileDependency}[1]{
  \typeout{(#1)}
  \@addtofilelist{#1}
  \IfFileExists{#1}{}{\typeout{No file #1.}}
}
\makeatother

\usepackage{setspace}
\usepackage[normalem]{ulem}
\useunder{\uline}{\ulined}{}
\usepackage{xparse}
\newsavebox{\fminipagebox}
\NewDocumentEnvironment{fminipage}{m O{\fboxsep}}
 {\par\kern#2\noindent\begin{lrbox}{\fminipagebox}
  \begin{minipage}{#1}\ignorespaces
 \end{minipage}\end{lrbox}%
  \makebox[#1]{%
    \kern\dimexpr-\fboxsep-\fboxrule\relax
    \fbox{\usebox{\fminipagebox}}%
    \kern\dimexpr-\fboxsep-\fboxrule\relax
  }\par\kern#2
 }

\title{How Much of the United States Can Still Host New Hyperscale Data Centers? A Constraint-Based Feasibility Analysis}

\author[*]{Milan Janosov}

\affil[*]{Geospatial Data Consulting Ltd.\\
\texttt{milan@janosov.com} \\
\href{https://www.janosov.com}{www.janosov.com}}

\begin{document}

\maketitle


\section*{Abstract}
{\small

The rapid expansion of hyperscale data centers, primarily driven by cloud computing and generative AI is placing growing pressure on electricity systems, land, and climate-sensitive infrastructure. While existing maps document where data centers are currently located, a major unanswered question remains: where can hyperscale data centers still be built under present-day physical, infrastructural, and environmental constraints?  

Here we address this question, focusing on the United States, using a national-scale, constraint-first geospatial framework that infers feasibility from revealed hyperscale siting patterns rather than from demand forecasts or optimization assumptions. By combining power-grid adjacency, environmental limits, land-use constraints, and climatic constraints within a uniform hexagonal spatial system, we estimate the feasible hyperscale hosting capacity.  

Our presented approaches converge on a limited feasible land envelope, implying a substantial contraction relative to naive land-availability assumptions. Based on observed build-out patterns, we estimate that total physically feasible U.S. hyperscale capacity lies in the tens of gigawatts rather than the hundreds. The results of this piece are intended to support national-scale reasoning about infrastructure feasibility under modern constraints.

}

\vspace{0.5cm}
{\small {\bf Keywords}: hyperscale data centers; AI infrastructure; GenAI; LLMs; spatial feasibility; geospatial analysis; power grid; climate risk; revealed preference; kernel density estimation; H3 hexagons}

\section{Introduction}

The rapid expansion of hyperscale data centers has become a defining feature of the modern digital economy. Growth in cloud computing, large language models, and generative AI has driven facilities operating at tens to hundreds of megawatts per campus from niche infrastructure into a material component of national electricity demand. National assessments indicate that data centers already account for a non-trivial share of U.S. electricity consumption \cite{shehabi2016united}.

Public reporting and open datasets have made the geography of existing data centers increasingly visible. Interactive maps and journalistic investigations, including those published by \emph{Business Insider} \cite{businessinsider_dc_locations}, highlight strong spatial concentration in a few regions, such as Northern Virginia and Dallas–Fort Worth. While these maps document where hyperscale infrastructure exists today, they do not answer a more fundamental question: where can hyperscale data centers still be built under present-day constraints?

This distinction is critical because hyperscale data centers are not freely locatable industrial assets in the U.S. Industry syntheses and policy analyses consistently emphasize that siting decisions are jointly constrained by access to bulk electricity transmission and generation, interconnection feasibility, cooling requirements, climate exposure, hydrological risk, terrain, protected lands, and urban and regulatory friction \cite{shehabi2016united,abdi2025hyperscale}. These constraints are discontinuous, spatially correlated, and slow to change, shaping both where hyperscale facilities cluster and where expansion is no longer viable.

Consistent with this, recent industry analyses show that hyperscale infrastructure is already highly concentrated, with roughly 60–62\% of global capacity located in just twenty metro or state-level markets \cite{synergy2024top20}. At the same time, industry guidance increasingly frames future siting as constrained primarily by power availability, grid interconnection, water access, climate risk, and regulatory friction rather than demand alone \cite{merkle2025siteselection,datacenterpower2025}.

Hence, in this work, we adopt a constraint-first geospatial perspective on hyperscale expansion. Rather than forecasting demand or optimizing investment, we ask a simpler question: under observed siting behavior and present-day physical conditions, what portion of U.S. territory remains physically and infrastructurally capable of hosting additional hyperscale capacity?

This framing contrasts with low-constraint environments—such as parts of Iceland or northern Scandinavia — where abundant power, cool climates, and limited land-use friction simplify siting decisions \cite{abdi2025hyperscale}. The contiguous United States represents the opposite case: a mature, heterogeneous, and increasingly constrained system in which feasibility must be inferred from how infrastructure has actually been deployed.

Unlike demand-driven forecasts, zoning-based suitability analyses, or parcel-level siting tools, this work estimates a national feasibility envelope conditioned on present-day infrastructure and revealed hyperscale siting behavior. The resulting capacity bounds should therefore be interpreted as physical and infrastructural ceilings under current patterns, not as projections of investment or construction. However, geospatial data science definitely possesses the toolset to explore those directions in the future.

We address this problem using a national-scale geospatial framework built on a uniform hexagonal grid and calibrated exclusively on revealed hyperscale behavior. Two unsupervised approaches - a gated similarity model and a kernel density–based feasibility surface—are used to estimate feasible land envelopes and translate them into bounded capacity estimates. We deliberately avoid demand forecasting, cost optimization, or assumptions about future grid expansion.

Our contribution is therefore not a prediction of how many data centers will be built, nor where investment should flow, but an estimate of how much hyperscale capacity the existing U.S. physical and infrastructural landscape can plausibly host under current patterns.


\section{Data}
\label{sec:data}

This work relies on nationally consistent, publicly available geospatial datasets covering the contiguous United States. The objective is to estimate large-scale physical and infrastructural feasibility for hyperscale data center siting, rather than parcel-level suitability, site optimization, or demand forecasting. In this section, first, we are going to overview the different types of information covered, and then list and briefly characterize each data source.

\paragraph{Power and grid infrastructure.}
Capturing the structural context of the electricity system is essential for assessing the large-scale feasibility of hyperscale data center siting. We represent this context using complementary datasets capturing bulk transmission adjacency, local interconnection proxies, and regional generation presence. High-voltage transmission lines are obtained from a national dataset published by the U.S. Fish and Wildlife Service \cite{usfws_transmission}. Electrical substations and transformers are extracted from OpenStreetMap \cite{osm_substations}. Given heterogeneous national attribute coverage, only geometric information is used, and no assumptions are made regarding voltage class, ownership, or available interconnection capacity. Power generation facilities and nameplate capacity are derived from the U.S. Energy Information Administration's EIA-860 dataset \cite{eia860}. To reduce noise from small generators, only plants with total nameplate capacity $\geq$50~MW are retained. Finally, the Electricity Market Module (EMM) regions are included for stratification and interpretation \cite{emm_regions}. 

\begin{figure}[hbt]
\centering
\includegraphics[width=1.0\textwidth]{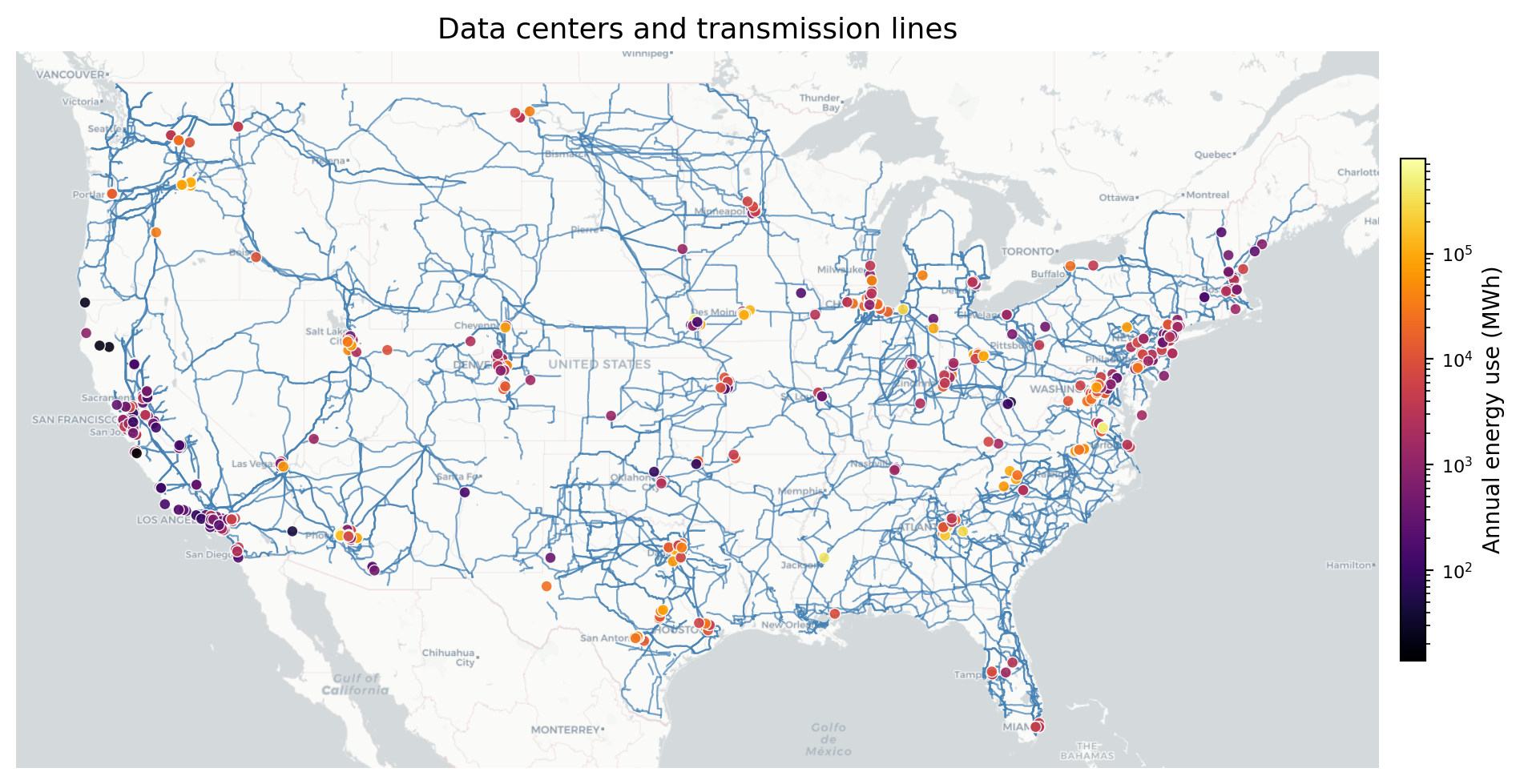}
\centering
\caption{USA-wide visualization of current data center locations and power transmission lines.}
\label{fig:dc_fig1}
\end{figure}

\paragraph{Climate, hydrology and flood risk.}
Cooling-related constraints are characterized using 30-year climate normals from the PRISM Climate Group \cite{prism}. As an example, we used the mean July temperature to capture sustained summer thermal load, and the maximum August temperature to represent peak thermal stress relevant for cooling system design and operational resilience. Surface water features are derived from the National Hydrography Dataset (NHD) \cite{nhd}. Wetlands are obtained from the National Wetlands Inventory \cite{nwi}, while flood exposure is represented using FEMA's National Flood Hazard Layer \cite{fema_flood}. 

\paragraph{Land, terrain, and built environment.}
Terrain constraints are derived from U.S. Geological Survey elevation data \cite{usgs_elevation}. Land-cover composition is represented using the National Land Cover Database \cite{nlcd}. Additionally, population density and built-up intensity are derived from the Global Human Settlement Layer \cite{ghsl} and serve as proxies for urban pressure and regulatory friction. Protected lands are identified using the Protected Areas Database of the United States \cite{padus}, and metropolitan context is represented using Core-Based Statistical Area (CBSA) boundaries \cite{cbsa}.

\paragraph{Existing data center locations}

Existing data center locations are compiled from a publicly available dataset published by Business Insider \cite{businessinsider_dc_locations}. These locations of the more than 1200 marked data centers are used exclusively to identify hyperscale reference environments and to validate spatial feasibility patterns. They are not used as predictors of future development. These data center locations, along with power transmission lines, are shown in Figure \ref{fig:dc_fig1}.

\paragraph{Data source summary.}

\begin{figure}[!hbt]
\centering
\includegraphics[width=0.85\textwidth]{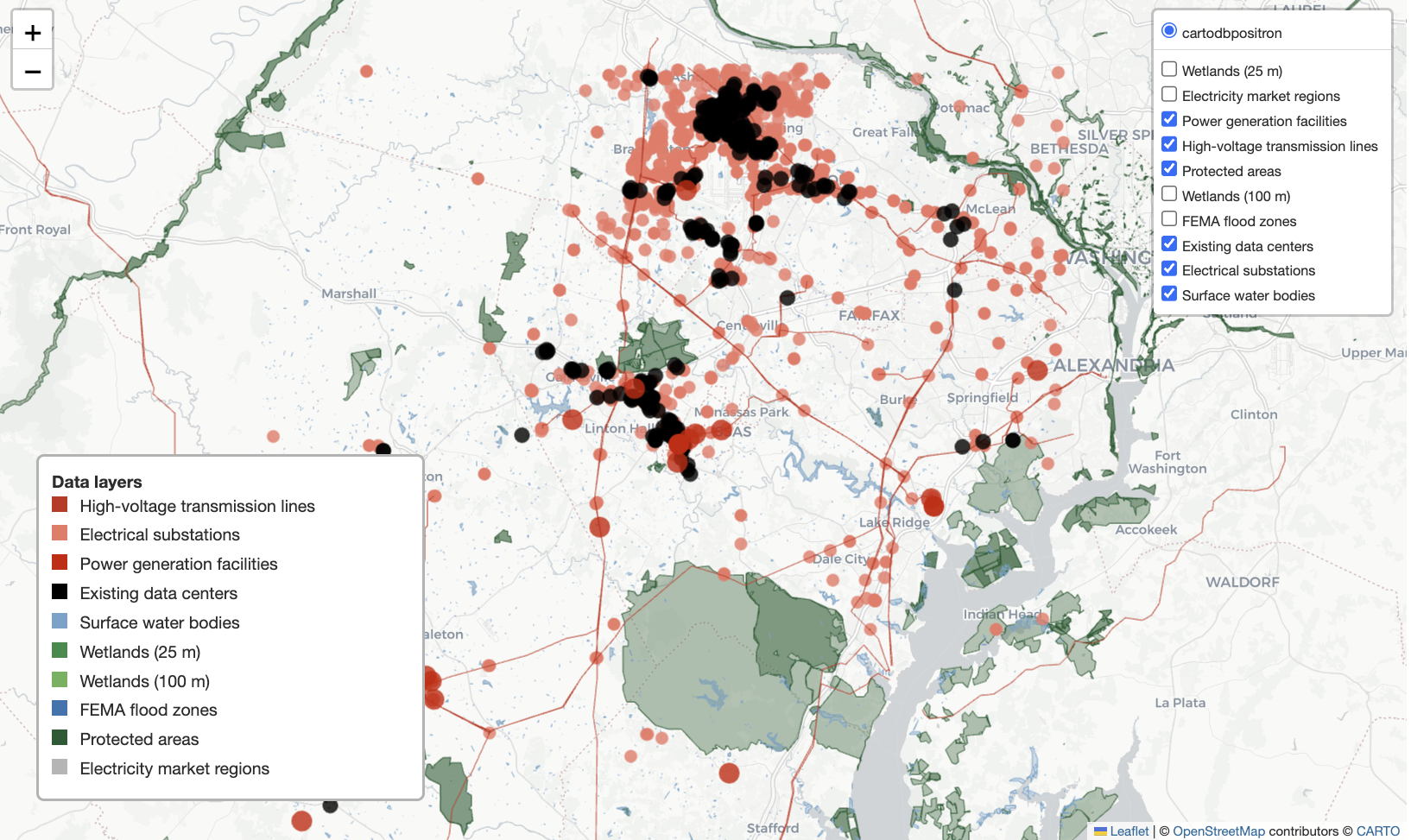}
\caption{Illustrative subset of the national data center dataset within the Northern Virginia hyperscale corridor. Points show existing data center locations scaled by upper-bound sustained power demand, while lines indicate high-voltage transmission infrastructure.}
\label{fig:dc_fig2}
\end{figure}

Below, Table \ref{tab:data_sources} lists all data sources we relied on, including their type, spatial resolution, and data source, while a close-up view of a few selected data sources are shown in Figure \ref{fig:dc_fig2} around the Northern Virginia hyperscale data center corridor.

\begin{table}[htbp]
\centering
\footnotesize
\caption{Summary of data sources used in the analysis. All datasets cover the contiguous United States.}
\label{tab:data_sources}
\begin{tabular}{p{4.5cm} p{2.5cm} p{3.0cm} p{4.0cm}}
\hline
\textbf{Dataset} & \textbf{Type} & \textbf{Resolution} & \textbf{Source} \\
\hline
Data center locations & Vector (points) & Facility-level & Business Insider \\
High-voltage transmission lines & Vector (lines) & Network scale & U.S. Fish and Wildlife Service \\
Substations & Vector (points) & Facility-level & OpenStreetMap \\
Power generation facilities & Vector (points) & Facility-level & U.S. EIA (EIA-860/923) \\
Electricity market regions & Vector (polygons) & Regional & U.S. DOE (EMM) \\
Elevation & Raster & $\sim$30 m & U.S. Geological Survey \\
Population density & Raster & 1 km & GHSL \\
Built-up area & Raster & 1 km & GHSL \\
Protected areas & Vector (polygons) & Parcel / area & PAD-US \\
Metropolitan areas (CBSA) & Vector (polygons) & Metro scale & U.S. Census Bureau \\
Surface water & Vector & Hydrographic & USGS NHD \\
Wetlands & Vector & Parcel / area & U.S. Fish and Wildlife Service \\
Flood hazard zones & Vector & Parcel / area & FEMA \\
Mean July temperature & Raster & 30 arc-sec & PRISM \\
Max August temperature & Raster & 30 arc-sec & PRISM \\
Land cover & Raster & 30 m & NLCD \\
\hline
\end{tabular}
\end{table}

\section{Methods}
\label{sec:methods}

\subsection{Spatial framework and resolution selection}
\label{sec:spatial_framework}

All analyses are conducted using Uber’s H3 hierarchical hexagonal grid, which offers a variety of spatial resolutions. To decide which hexagon size and resolution we relied on, we followed a modeling protocol. Namely, density-based clustering (DBSCAN) applied to known data center locations. This clustering yields stable clustering radii of 23–27 km across data center capacity thresholds, corresponding to cluster areas of approximately 40–100 km$^2$. These scales align reasonably well with H3 resolution 4 (hexagon area $\approx$25 km$^2$), which is adopted as the canonical analysis resolution throughout the following analytical steps. At H3 resolution 4 (4,348 hexagons nationally), 40 hexagons (0.9\%) are classified as hyperscale, 151 (3.5\%) as non-hyperscale data center regions, and 3 (<0.1\%) remain uncertain, illustrating the narrow spatial footprint of revealed hyperscale environments.

\subsection{Feature construction and aggregation}
\label{sec:feature_construction}

All datasets were projected to a common metric coordinate reference system and aggregated to a national H3 hexagonal grid at resolution 4. Features were constructed as interpretable, monotonic indicators of physical, infrastructural, environmental, and contextual constraints rather than as direct predictors of development outcomes.

\begin{figure}[H]
\centering
\includegraphics[width=\textwidth]{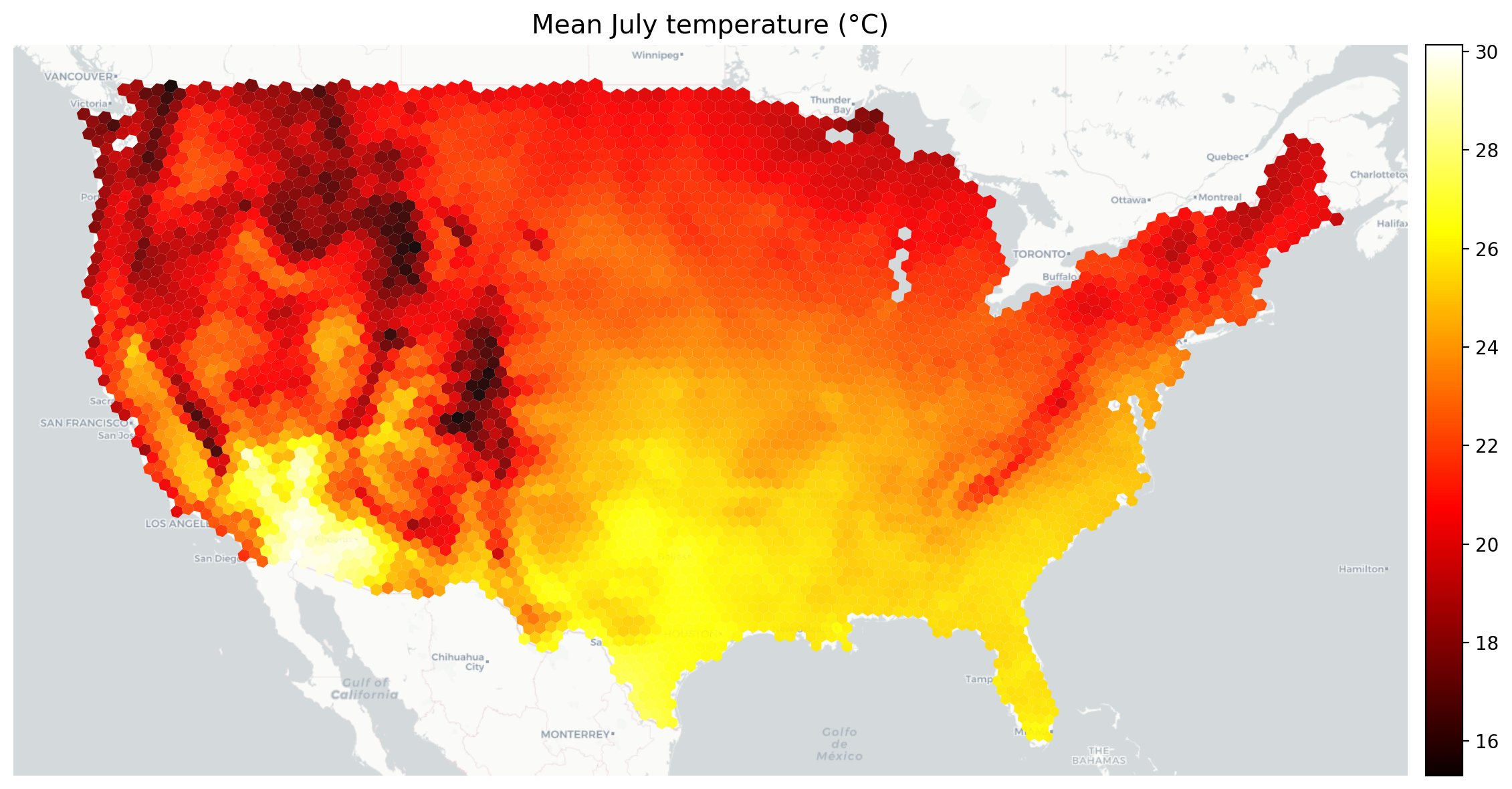}
\caption{Mean July temperature aggregated to H3 resolution 4 hexagons.}
\label{fig:dc_fig3a}
\end{figure}

\begin{figure}[H]
\centering
\includegraphics[width=\textwidth]{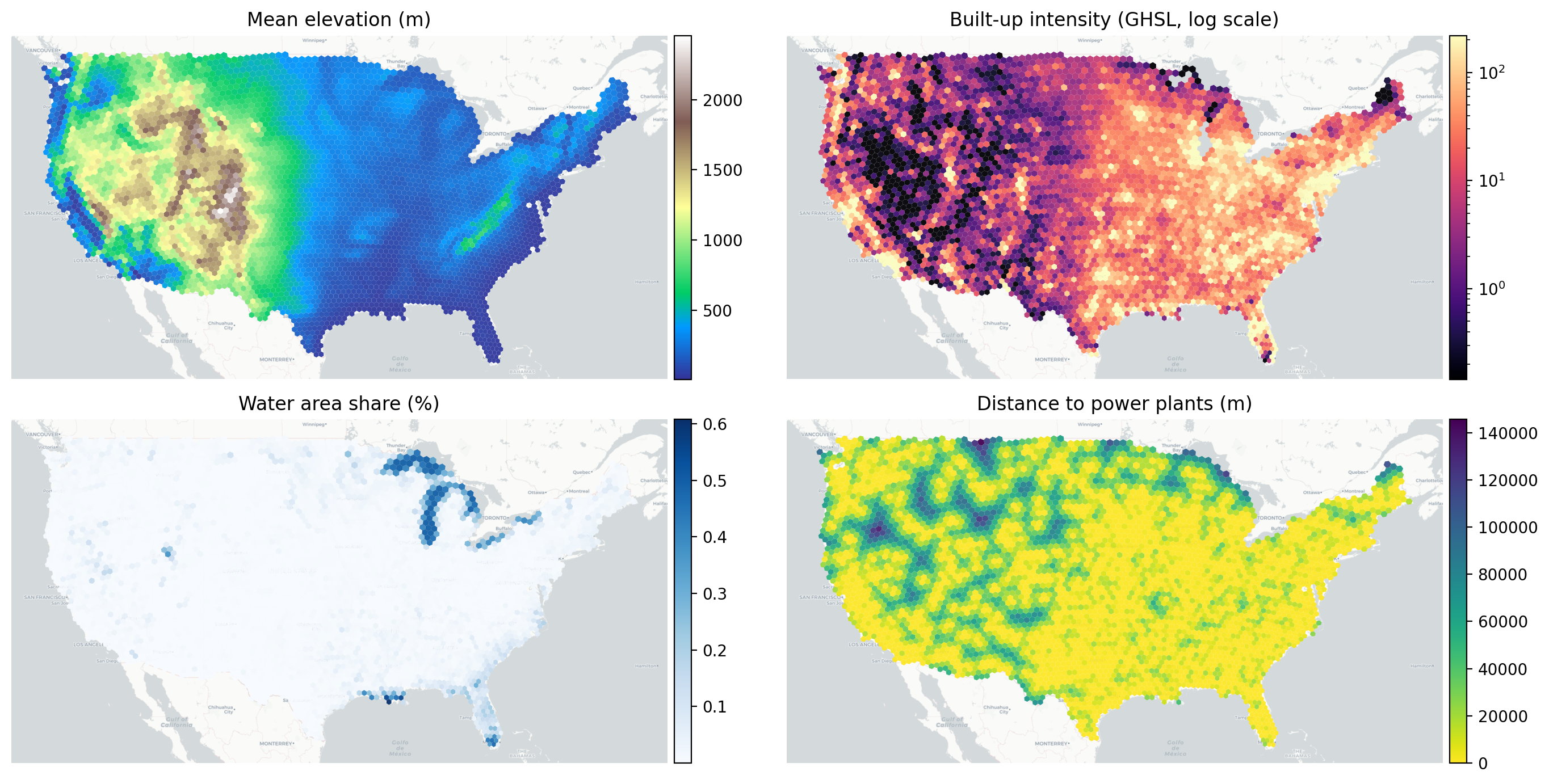}
\caption{Environmental and infrastructural context layers: elevation, built-up intensity (log scale), surface water share, and distance to power generation facilities.}
\label{fig:dc_fig3b}
\end{figure}

\paragraph{Climate, hydrology, and flood risk.}
Cooling-related environmental constraints were represented using climate normals and hydrological exposure indicators. Mean July temperature captures sustained summer thermal load, while maximum August temperature represents peak heat stress (Figure~\ref{fig:dc_fig3a}). Hydrological constraints were quantified as fractional area overlaps per hexagon, including surface water coverage, wetland exposure, and flood hazard exposure (Figure~\ref{fig:dc_fig3b}).

\paragraph{Power and grid infrastructure.}
The bulk electricity system context was captured using a set of structural adjacency and regional context indicators. Transmission accessibility was quantified using two complementary features: the minimum Euclidean distance from each hexagon polygon to the nearest high-voltage transmission line ($\geq$115~kV) and the total length of such lines intersecting the hexagon. Local interconnection plausibility was proxied using OpenStreetMap substations, computing both the minimum distance to the nearest substation and the count of substations within each hexagon. Regional generation context was characterized using large power plants ($\geq$50~MW), computing distance to the nearest plant and total nameplate capacity within a 100~km radius of the hexagon centroid (Figure~\ref{fig:dc_fig3b}). Dominant generation fuel within the same radius and Electricity Market Module (EMM) region assignment were retained for stratification and interpretation only. All power-related features represent structural context rather than deliverability, capacity availability, pricing, or regulatory feasibility.

\paragraph{Land, terrain, and built environment.}
Topographic constraints were derived from elevation data, computing per-hexagon statistics including mean elevation, elevation variability and range, mean slope, upper-tail slope, and the share of steep terrain. Urban pressure and regulatory friction were proxied using GHSL-derived population and built-environment indicators (Figure~\ref{fig:dc_fig3b}), including total population, upper-tail population density, log-transformed population, and built-up intensity. Land-cover composition was summarized using NLCD-derived developed and open land shares as contextual descriptors. Protected lands were applied as hard exclusions using binary feasibility masks and fractional area overlap. Metropolitan context was captured using CBSA membership and identifiers for stratification and interpretation rather than suitability scoring (Figure~\ref{fig:dc_fig3b}).

\subsection{Identifying hyperscale reference regions}
\label{sec:hyperscale_identification}

We constructed a national-scale classification of hyperscale and non-hyperscale data center regions using a conservative, constraint-first spatial framework. Known data center locations are aggregated into Uber H3 hexagonal cells across multiple spatial resolutions (r2–r5), with resolution 4 adopted as the primary analytical scale based on empirical clustering behavior.

For each hexagon, lower- and upper-bound estimates of sustained average power demand (MW) are derived from reported annual energy consumption. Hexagons are assigned high-confidence labels using conservative thresholds: cells with aggregate upper-bound demand of at least 20~MW are labeled hyperscale, while cells with aggregate lower-bound demand of at most 5~MW are labeled non-hyperscale. Remaining cells are treated as uncertain.

To resolve uncertain cases without introducing circularity, we train a binary classifier using only high-confidence hyperscale and non-hyperscale hexagons. The classifier relies exclusively on power infrastructure and grid context features, including proximity to transmission lines, substations, power plants, and regional generation capacity. Power demand estimates and environmental variables are explicitly excluded. Balanced subsampling and repeated training are used to quantify prediction uncertainty, and final labels are assigned using an uncertainty-aware decision rule based on the mean and variance of predicted hyperscale probability.

At H3 resolution 4, this procedure identifies a small and spatially clustered set of hyperscale hexagons. Following classification, 40 hexagons are flagged as hyperscale, 151 as non-hyperscale data centers, and only 3 remain uncertain. Aggregating upper-bound sustained demand across the hyperscale regions yields approximately 4.5–5.0 GW of current hyperscale load, corresponding to roughly 40–45 TWh per year, or about ~1\% of total U.S. annual electricity consumption (roughly 4,000 TWh). This estimate is stable across the spatial resolutions used in the analysis and should be interpreted as a conservative lower bound, reflecting publicly observable hyperscale facilities rather than a comprehensive census.

\subsection{Similarity-based feasibility analysis}
\label{sec:similarity_model}

First, we assess spatial feasibility using a similarity-based, unsupervised model grounded in revealed hyperscale behavior. Each hexagonal cell is characterized using a comprehensive set of environmental, infrastructural, and contextual indicators introduced earlier.

To enable robust comparison across heterogeneous variables, all numeric features are transformed using rank-based uniform scaling. Hyperscale hexagons are used exclusively to define reference envelopes in feature space, summarized using empirical quantiles. 

Similarity between candidate locations and the hyperscale envelope is computed using a distance-based metric with an exponential kernel, yielding a continuous similarity score. Crucially, similarity is decomposed into environmental and power components and combined non-compensatorily by taking their minimum. This ensures that favorable environmental conditions cannot offset inadequate electricity infrastructure.

Candidate regions are identified using conservative similarity thresholds derived from the hyperscale distribution itself, including the minimum, 5th percentile, and 10th percentile similarity values. These thresholds define nested feasibility envelopes rather than binary classifications.

To translate spatial feasibility into aggregate capacity, we estimate hyperscale power density from observed facilities and apply conservative (25th percentile), typical (median), and upper (75th percentile) densities to the total area of candidate regions. Adding current hyperscale capacity yields a conditional national envelope of approximately 24–78~GW under present-day siting patterns. These values represent physical and infrastructural upper bounds rather than forecasts.

\subsection{Kernel density–based feasibility analysis}
\label{sec:kde_model}

As a robustness check, we estimate a kernel density–based feasibility surface calibrated exclusively on revealed hyperscale behavior. Using the same H3 resolution and feature set, all numeric features are transformed to rank-based uniform scores.

Then, two independent kernel density models are fitted using hyperscale hexagons only: one describing power-system feasibility and one describing environmental and land-context feasibility. For each hexagon, likelihood scores under both models are computed and combined non-compensatorily by taking their minimum, enforcing a strict power-gated feasibility logic.

Candidate hyperscale locations are defined using distribution-derived thresholds of the combined KDE score, producing conservative and permissive feasibility envelopes. No supervised learning, optimization, or demand assumptions are used at any stage.

Feasible land area is translated into a bounded capacity estimate by applying the empirical distribution of observed hyperscale power density to each feasibility envelope and adding current hyperscale capacity. This yields a likely national capacity range of approximately 26–106~GW, with conservative assumptions producing lower bounds closely aligned with the similarity-based model.

\subsection{Methodology note}

The two feasibility surfaces presented in this paper are not intended to be independent estimators in a statistical or predictive sense. Both are deliberately constructed from the same underlying constraint primitives, reflecting the fact that hyperscale siting decisions are governed by a limited and shared set of physical, infrastructural, and environmental factors. The purpose of using two methods is therefore not cross-validation, but robustness: to demonstrate that national-scale feasibility patterns remain stable across different aggregation logics applied to the same revealed-preference constraints.

This study does not pursue internal validation in the predictive or classification sense (e.g., accuracy, AUC, or clustering goodness metrics), because it does not aim to predict new data center locations. Instead, known hyperscale locations are treated as revealed-preference reference environments that define feasible conditions. The objective is to infer the extent of national land that remains structurally similar to environments where hyperscale facilities have already been built. Robustness is assessed through stability across spatial resolutions and across the two aggregation frameworks.

\section{Results}
\label{sec:results}

Hyperscale data centers are highly spatially concentrated, with approximately 40 hexagons classified as hyperscale at H3 resolution 4. Aggregated upper-bound sustained load across these hyperscale hexagons is approximately 4.5--5.0~GW, corresponding to roughly 40--45~TWh/yr of electricity consumption as of today.

\begin{figure}[!hbt]
\centering
\includegraphics[width=0.6\textwidth]{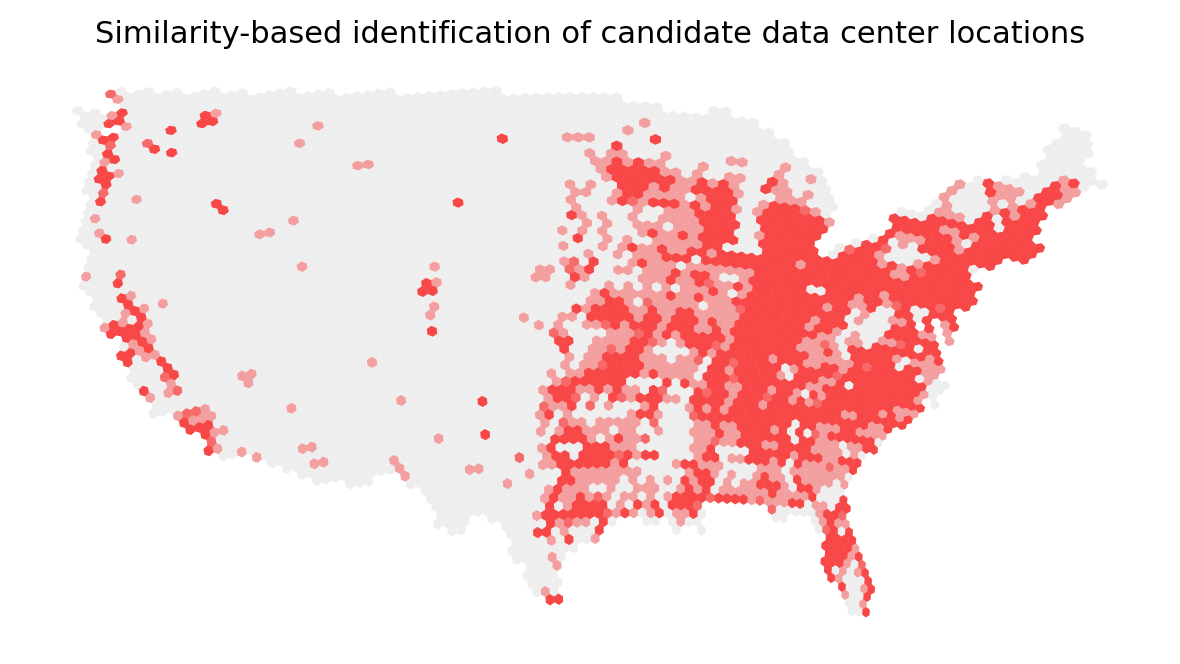}

\vspace{0.4cm}

\includegraphics[width=0.6\textwidth]{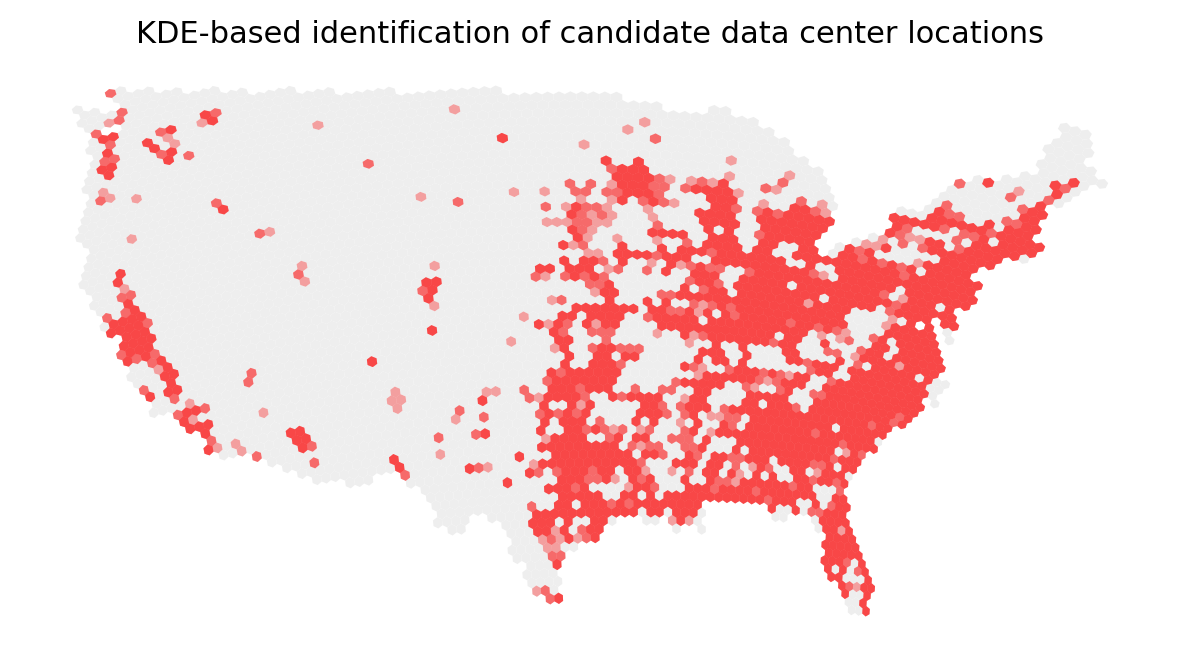}
\caption{Comparison of feasibility envelopes derived from (top) similarity-based and (bottom) KDE-based approaches. Darker regions indicate hexagons selected under multiple feasibility thresholds.}
\label{fig:dc_fig4}
\end{figure}

Both the similarity-based and KDE-based approaches identify a limited but spatially coherent feasible land envelope for additional hyperscale development based on the wide range of constraints included (Figure~\ref{fig:dc_fig4}). Translating these envelopes using empirically observed hyperscale power densities yields conservative national capacity ranges of approximately 24--78~GW of total physically feasible hyperscale capacity.

{\bf The two unsupervised models converge on a consistent estimate, with lower-to-mid feasible capacity ranges centered around 25-50~GW}. The close agreement between the approaches indicates that these bounds are imposed primarily by structural constraints---including existing electricity infrastructure, environmental context, and land-use patterns---rather than by modeling assumptions or methodological choices.

Feasible hexagons may include dense metropolitan regions; in such cases, feasibility reflects regional electricity infrastructure and revealed hyperscale clustering rather than parcel-level buildability, with development implicitly occurring in peripheral, industrial, or previously established data-center subareas.

All estimates are intentionally conservative and should be interpreted as feasibility envelopes rather than forecasts. The upper end of the estimated range represents a physical ceiling under present-day siting patterns and infrastructure conditions, implying that substantially exceeding 100~GW of hyperscale capacity would require significant transformation of the underlying energy system, regulatory environment, or data center siting practices.

\section*{Declarations}

\paragraph{Funding.}  This research received no external funding.

\paragraph{Use of AI tools.}  Generative AI tools were used to assist with language refinement and code development.

\section*{Acknowledgments}

I am grateful to my friend and colleague, Manran Zhu, for her thoughtful and forward-looking suggestions and insightful feedback throughout the development of this work.

\end{document}